\documentclass[]{rmaa}

\usepackage{paralist}

\usepackage{psfrag,color}
\usepackage{graphicx}
\usepackage[latin1]{inputenc}

\title{Discovery of New Faint Northern Galactic Planetary Nebulae} 

\author{
  Agn\`es Acker,\altaffilmark{1} 
   Henri M. J. Boffin,\altaffilmark{2}
  Nicolas Outters,\altaffilmark{3}
  Brent Miszalski,\altaffilmark{4,5}
   Laurence Sabin,\altaffilmark{6}
    Pascal Le D\^u,\altaffilmark{7}
  and Filipe Alves\altaffilmark{8}}

\altaffiltext{1}{Observatoire de Strasbourg, France.}
\altaffiltext{2}{ESO, Chile.}
\altaffiltext{3}{Orange Observatory, France.}
\altaffiltext{4}{South African Astronomical Observatory.}
\altaffiltext{5}{Southern African Large Telescope Foundation.}
\altaffiltext{6}{Instituto de Astronom\'{\i}a, Universidad Nacional Aut\'onoma de M\'exico.}
\altaffiltext{7}{Porspoder, France.}
\altaffiltext{8}{CROW-Portalegre, Atalaia group, Portugal.}

\shortauthor{Acker et al.}
\shorttitle{New faint Planetary Nebulae discovered}

\fulladdresses{
\item Agn\`es Acker:  Observatoire de Strasbourg, 11 rue de l'universit\'e,  F-67000 Strasbourg, France (agnes.acker@astro.unistra.fr).
\item Henri M. J. Boffin: ESO, Alonso de Cordova 3107, Vitacura, Casilla 19001, Santiago, Chile
(hboffin@eso.org).
\item Nicolas Outters: Orange Observatory, 50 impasse des Gorges,  F-74800 Saint-Sixt, France
(nico.outters@libertysurf.fr).
\item Brent Miszalski: South African Astronomical Observatory and Southern African Large Telescope Foundation, PO Box 9, Observatory 7935, South Africa (brent@saao.ac.za).
\item Laurence Sabin: Instituto de Astronom\'{\i}a, Universidad Nacional Aut\'onoma de M\'exico,
Apdo. Postal 877, 22800 Ensenada, B.C., Mexico
(lsabin@astrosen.unam.mx).
\item Pascal Le D\^u: 21 route de Kermerrien,  F-29840 Porspoder, France (ledu@shom.fr).
\item Filipe Alves: CROW-Portalegre, Atalaia group, Portugal
(manalokos@gmail.com).
}

\listofauthors{A. Acker et al.}
\indexauthor{Acker, A.}
\indexauthor{Boffin, H.M.J.}
\indexauthor{Outters, N.}
\indexauthor{Miszalski, B.}
\indexauthor{Sabin, L.}
\indexauthor{ Le D\^u, P.}
\indexauthor{Alves, F.}

\abstract{
We present the discovery of 6 nebular objects made by amateur astronomers. Four of these discoveries are clearly planetary nebulae (PNe), one is a possible PN, and another is a likely H~II region. 
The bipolar nebula Ou4 presents the largest angular extent ever found : over one degree on the sky! We consider various scenarios that could explain such a nebula. Ou4 could be one of the nearest PNe known, though its possible PN nature will need confirmation.
 }

\resumen{}

\addkeyword{ISM: planetary nebulae}
\addkeyword{Surveys}

\begin{document}
\maketitle

     \section{Introduction}

According to the most generally accepted paradigm, planetary nebulae (PNe) represent the luminous transitory phase of the end of the life of low and intermediate mass stars (0.8 to 8 M$_\odot$). About 2,850 PNe have been detected so far in the Galaxy, according to the new consolidated database of PNe  \citep{Mi2012}, which gather the PNe listed in the Strasbourg catalogue \citep{SECGPN}, the PNe discovered in the Southern Hemisphere by the AAO-UKST H$\alpha$ Surveys (MASH;  \citet{Parker}, \citet{Mi2008}), and in the northern IPHAS Survey \citep[Sabin et al., in prep.]{Drew,Vii2009}. 

However, the expected Galactic PN population is estimated to be between 6,000 and 25,000 or more \citep{Frew06,Jacoby10}, depending on the possible role of binaries in forming PNe. 
\citet{Moe2006} argued, based on population synthesis, that only $\sim$20\% of low- and intermediate-mass stars produce a PN, with the remainder transiting between the AGB and white dwarf phases with invisible, or under-luminous nebulae. \citet{SS2005} predicted that deep searches could find the brightest among these underluminous PNe and that they would be spherical. This prediction has only been partly borne out by the MASH surveys mentioned above that doubled the fraction of spherical PNe from $\sim$10\%  to $\sim$20\% and by the Deep Sky Hunters survey that found a similar fraction  \citep{Jacoby10}. It is thus necessary to try to find many more faint PNe, so as to clearly establish the necessary observational framework to compare with theory. 
 
 In this paper, 
we present the detection of 4 new PNe and 1 possible new faint PN from deep multi-colour images performed by amateur-astronomers, in addition to one H~II region as an example of a ``tricky'' object. The paper is organised as follows: in Sect.~2, we present the imaging programmes that led to the discoveries, while Sect.~3 presents spectroscopy for two of the brightest of the newly found objects. Finally, Sect.~4 looks into more detail into each object.
   
    \section{Search methods}

As mentioned above, a group of  amateur astronomers named the Deep Sky Hunters \citep{K2006} have been methodically scanning the digital sky survey (DSS), thereby visually identifying dozens of faint PN candidates in the Galactic Plane. Most objects were confirmed by spectroscopy or through narrow-band CCD imaging. The discovery of 60 faint PNe using this method is reported in \citet{Jacoby10}. This sample is notable for the large proportion of spherical PNe at high latitudes, a trait first seen in MASH-I and MASH-II PNe (Parker et al. 2006; Miszalski et al. 2008). 

Visual searches of all-sky plates are much less sensitive to compact PNe which may be an important complement at high latitudes where distant Halo PNe are expected to reside. \citet{Mi2011} introduced the Extremely Turquoise Halo Object Survey (ETHOS) and its first discovery, ETHOS~1 (PN G068.1$+$11.0), which is a compact PN that hosts a pair of jets and a close binary central star ($P=0.535$ d). The ETHOS takes a similar approach to MASH-II by processing the catalogue photometry from Hambly et al. (2001) in search of new PN candidates and thereby explores a complementary parameter space to \citet{Jacoby10}.

There is, however, another way in which amateurs can help in discovering new, faint PNe. Since about 10 years ago, amateur astronomers have been equipped with CCD imagers with high sensitivity and $\sim$1 arcsec resolution, allowing them to obtain images with high resolution and contrast, and with unprecedented depth using narrow-band filters. The most impressive example so far is without doubt the ``Soap Bubble'' that measures $\sim$260\arcsec\ across and features a $B_J=19.45$ mag central star as measured by the SuperCOSMOS catalogue. Discovered by Dave  \citet{Jura2010}, it was then painstakingly confirmed as being a planetary nebula (Ju~1, PN G075.5+01.7; J2000 20$^h$15$^m$22$^s$.2, +38$^{\circ}$02\arcmin58\arcsec). It is noteworthy that, relating his discovery in a popular magazine, Jurasevich recognised the difficulty of reporting it in an IAU Central Bureau Electronic Telegram, which is not really equipped to handle discoveries of objects of a non-transient nature such as PNe. 

Here we report on the use of similar deep narrow-band imaging to find new PNe that was performed by amateur astronomers F. Alves, N. Outters and P. Le D\^u in the vicinity of various H~II regions (Sharpless 1953) and PN G080.3$-$10.4 (MWP1; Motch, Werner \& Pakull 1993). The high sensitivity, exclusive telescope time and large field of view of their setup greatly contributed to their discoveries.

To confirm their discoveries, we cross-checked the candidate objects on POSS-I and POSS-II images to guard against plate defects \citep{Parker}. 
Then, we checked whether the object was not already cataloged (e.g. in SIMBAD or in the PN catalogues of \citet{SECGPN,Parker,Mi2008}).
Additional digitised photographic $B$, $R$ and $I$ images (Hambly et al. 2001) and H$\alpha$ images from the IPHAS survey (Drew et al. 2005; Gonz\'alez-Solares et al. 2008) were then retrieved. The former allow for faint blue central stars to be identified, while the latter allow the nebula morphology to be assessed.

\begin{table*}
 \caption{Characteristics of the newly discovered nebulae.}
   \label{tab:booktabs1}
\centering
 \begin{tabular}{lllllll}\toprule
{\bf OBJECT}	& {\bf Ou1} &  {\bf Ou2}  & {\bf Ou3}	\\
\midrule
PN G  	          & Possible PN? or	&120.4-01.3	&059.2+01.0	\\
			  &  small H~II region?	&			& 			\\
\hline
Position J2000	&04 07 21.6
&00 30 56.8 
&19 38 17.5 
 \\
& +51 24 22	& +61 24 34	& +23 45 49	\\
\hline
Size 	&75\arcsec x 55\arcsec	& 80\arcsec	& 92\arcsec 	\\
\hline
CSPN& $B_J$ = 19.38& $B_J$ = 19.58 & $B_J \sim$21 \\
magnitude	&$B_J - R_F = -0.60$  & $B_J - R_F = -0.26$ & &\\
\hline
Discoverer& Outters N. &  Outters N. &  Outters N. \\
Discovery date	&
2007-2009	& 
Nov 2010	& 
Nov. 2009	 \\
\hline
Instrument Optic & FSQ106 F/5  &FSQ106 F/5 & FSQ106 F/5 \\
CCD camera	& 
FLI img6303e & 
SbigSTL6303e	& 
FLI img6303e\\
\hline
Exposure time	& 9h H$\alpha$ & 20.5h H$\alpha$   & 11h H$\alpha$\\
& 9.5h S[II] & 
5.6h BVR	& 
11.5h S[II] \\
& 5.5h O[III]	& & 13.5h O[III]\\
\hline
Spectrum (exp. time) & -- & 40 min & -- \\
       date 	& -- & 	
23-26/08/11	& --\\
\hline
Other images&	--	& IPHAS& 	IPHAS\\
& &  WISE W4 (22 $\mu$m)& \\
& & POSS2/UKSTU& \\
\hline
Object nearby	& NGC 1491	&Sh2-173	& NGC 6820\\
\bottomrule
 \end{tabular}
 \end{table*}
 
\begin{table*}
 \caption{Characteristics of the newly discovered nebular objects (Cont.).}
   \label{tab:booktabs2}
\centering
 \begin{tabular}{lllllll}
\toprule
{\bf OBJECT}		& {\bf Ou4}  & {\bf lD\^u1} &	{\bf Alv1}\\
\midrule
PN G  	          &098.5+07.9 (TBC)	&094.5-00.8a	&079.8-10.2\\
\hline
Position J2000	
&21 11 48.2 
&21 36 05.5 
&21 15 06.6 \\
& +59 59 12	  & +50 54 10	& +33 58 18\\
\hline
Size 	&1$^{\circ}$9\arcmin24\arcsec $\times$ 19\arcmin48\arcsec	& 132\arcsec 	& 270\arcsec\\
\hline
CSPN&unclear  & $B_J$ = 21.64& $B_J$ = 18.22\\
magnitude	& (see text) & & $B_J - R_F = -0.40$\\
\hline
Discoverer&  Outters N. & Le D\^u P. &  Alves F.\\
Discovery date	&
June 2011 &
Aug. 2011	 &
Nov. 2009 \\
\hline
Instrument Optic & FSQ106 F/5 & FSQ106 F/5 & 8\arcsec\ ASA astrograph\\
CCD camera	& 
SbigSTL6303e 	& 
QSI 583wsg	&  F/3.7-FLI Microline\\
& & KAF8300 & KAF8300 \\
\hline
Exposure time	& 18h H$\alpha$ & 18h20 H$\alpha$  & 14.1h H$\alpha$\\
&  8h S[II] &13h40 S[II]  & 8.6h O[III]\\
& 12.5h O[III]  & 11h40 O[III]	 & BVR\\
\hline
Spectrum (exp. time) & 40 min & -- & -- \\
       date 	& 
23-26/08/11	& --	& --\\
\hline
Other images&	--	&IPHAS& POSS-II DSS red\\
 & &
   P\"opsel \& Binnewies	& \\
\hline
Object nearby	& Sh2-129 & Sh2-124 & 	MWP1\\
 & HR 8119	& PN G094.5-00.8 & \\
\bottomrule
 \end{tabular}
 \end{table*}

Tables~\ref{tab:booktabs1} and ~\ref{tab:booktabs2} present the characteristics of the 6 newly discovered objects. The following information is provided:
 \begin{itemize}
\item	lines 1 and 2: the usual name (truncated author's name) and the PN G denomination
\item	lines 3 and 4: the J2000 coordinates, and the angular size
\item	line 5: the CSPN identification and magnitude, if available. The $B_J$ and $R_F$ magnitudes correspond to the ones determined by the Second Palomar Sky Survey \citep{1991PASP..103..661R} made available by Hambly et al. (2001)
\item	line 6: the name of the discoverer, and the discovery date
\item	lines 7 and 8: the technical details concerning the telescope and CCD camera, and the exposure times using different filters.
\item	line 9: the exposure time and date of the spectroscopy, if any
\item	line 10: details on other images of the object available
\item	line 11: the name of objects observed near the newly discovered object.
\end{itemize}
 
The observations were done using a 200mm F/D 3.7 newtonian telescope (Alves), and a Takahashi FSQ106 - F/D 5  telescope (Le D\^u and Outters), with commercially available cooled CCD cameras (Fingerlake and Sbig with KAF6303e, QSI and Fingerlake with KAF8300). 
The narrow-band filters are centred on 656 nm (H$\alpha$), 672.5 nm ([SII]), and 500.7 nm  ([OIII]), respectively. 
Alves used filters of the brand Astronomik and whose width were 6 nm for H$\alpha$ and 12 nm for [OIII], while Le D\^u and Outters used filters from Astrodon. These have a FWHM of 5 nm, except the [OIII] filter used to observe Ou4 that had a FWHM of 3 nm.
The various narrow-band images were digitally combined, so as to produce colour images.

 \begin{figure*}[hbtp]
  \begin{center}
     \includegraphics[scale=0.205,angle=0]{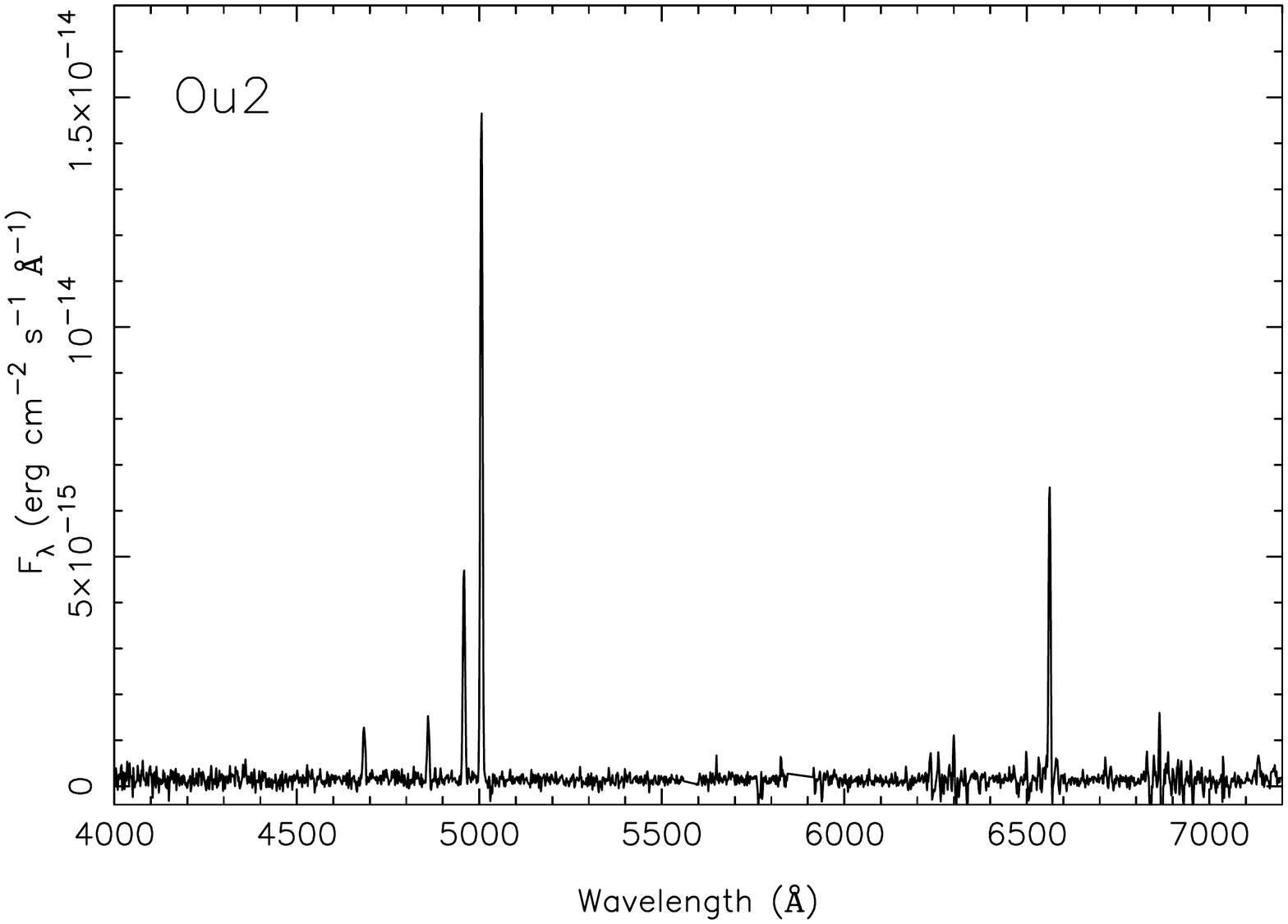}
     \includegraphics[scale=0.205,angle=0]{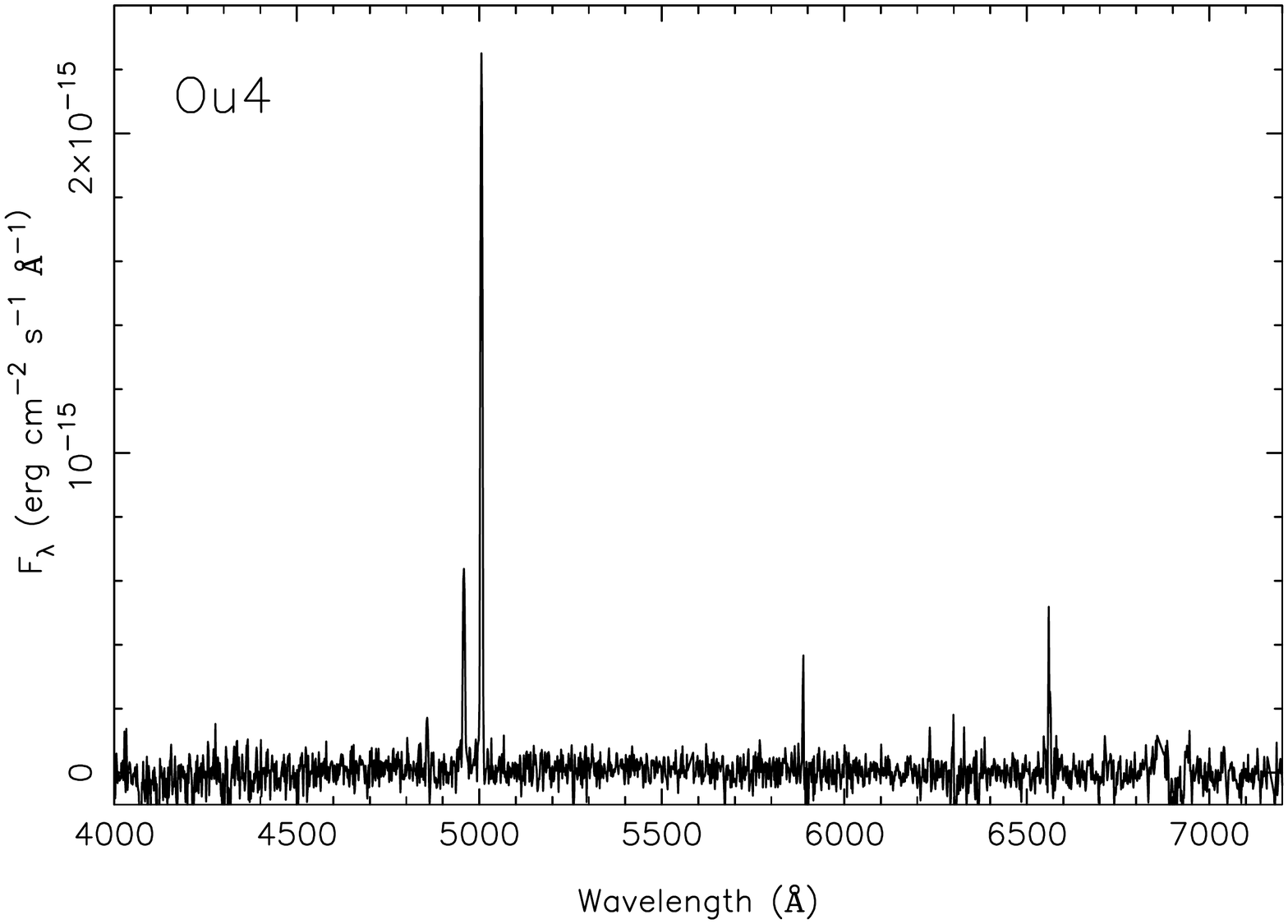}
  \caption{SPM spectra of Ou2 (left) and Ou4 (right). 
  }
  \label{fig2} \end{center}
\end{figure*}

 \section{Spectroscopy}
 
We obtained spectra for the two brightest of the six candidates, i.e. Ou2 and Ou4. 
The observations were performed by one of us with the 2.1-m telescope at the San Pedro M\'artir Observatory (OAN-SPM\footnote{The Observatorio Astron\'omico Nacional at the Sierra de San Pedro M\'artir (OAN-SPM) is a national facility operated by the Instituto de Astronom\'\i a of the Universidad Nacional Aut\'onoma de M\'exico.}) equipped with a Boller \& Chivens spectrograph and a Marconi 2 CCD (2048 $\times$ 2048 pixels with a spatial resolution of 1.22\arcsec/pixel). The long slit had a length of 5\arcmin\  and a width of 200 $\mu$m (about 2\arcsec).
A position angle of 90 degrees was adopted as well as an exposure time of 2400 seconds. Due to the faintness of the targets we also took great care while positioning the slit, making sure it also contains enough sky for background subtraction. For Ou4, the slit was placed on the brightest part of the nebula, i.e. on the eastern edge of the southern lobe. 
The observations took place between 23--26 August 2011. A 400 l/mm grating was used in order to cover the range $\sim$4330--7530 {\AA} with a resolution of $\sim$5 \AA (FWHM). We used a 2x2 binning to increase the signal. The observing conditions were variable during the run with thin meandering clouds and therefore non-photometric sky.
The data were bias subtracted, flatfielded, and wavelength calibrated using a copper-argon lamp. The flux calibration was done using the spectrophotometric standard star BD+284211, which was observed with a wide slit of 1200 $\mu$m. The reduction was performed using standard routines in IRAF.

\begin{table}
\caption{Emission line intensities (uncorrected for interstellar reddening) of Ou2 and Ou4.   }
   \label{tab:spectro}
   \begin{center}
   \begin{tabular}{lrcr}\toprule
{\bf Line} 		& 	{\bf Ou2} 	{\bf (H$\beta$=100)} 	& 	& 		{\bf Ou4} 	{\bf (H$\beta$=100)}	\\\midrule
He II 4686		& 	100				& 	& 					\\	
H$\beta$ 4861	&       100				& 	&  	100$\colon$	\\
$[$O III$]$ 4959		& 362				& 	& 	462				\\
$[$O III$]$ 5007		& 	1153				& 	& 	1533				\\
H$\alpha$ 6563& 	453				& 	& 	227$\colon$		\\
\hline
$[$N II$]$ 6583		&      trace				&	& 					\\
$[$Ar III$]$ 7135	&	trace				& 	& 					\\
\bottomrule
  \end{tabular}\\
{\small For Ou4, the H$\beta$ detection is very uncertain and the quoted\\ value is only an upper limit.
The H$\alpha$/H$\beta$ line ratio leads\\ to a visual extinction constant, 
c$_{H\beta}$ = 2.84~$\log$((H$\alpha$/H$\beta$)/2.86),\\
equal to 0.6 for Ou2 and about 0 for Ou4.} \end{center} 
 \end{table}

\begin{figure*}
  \begin{center}
     \includegraphics[scale=0.475,angle=0]{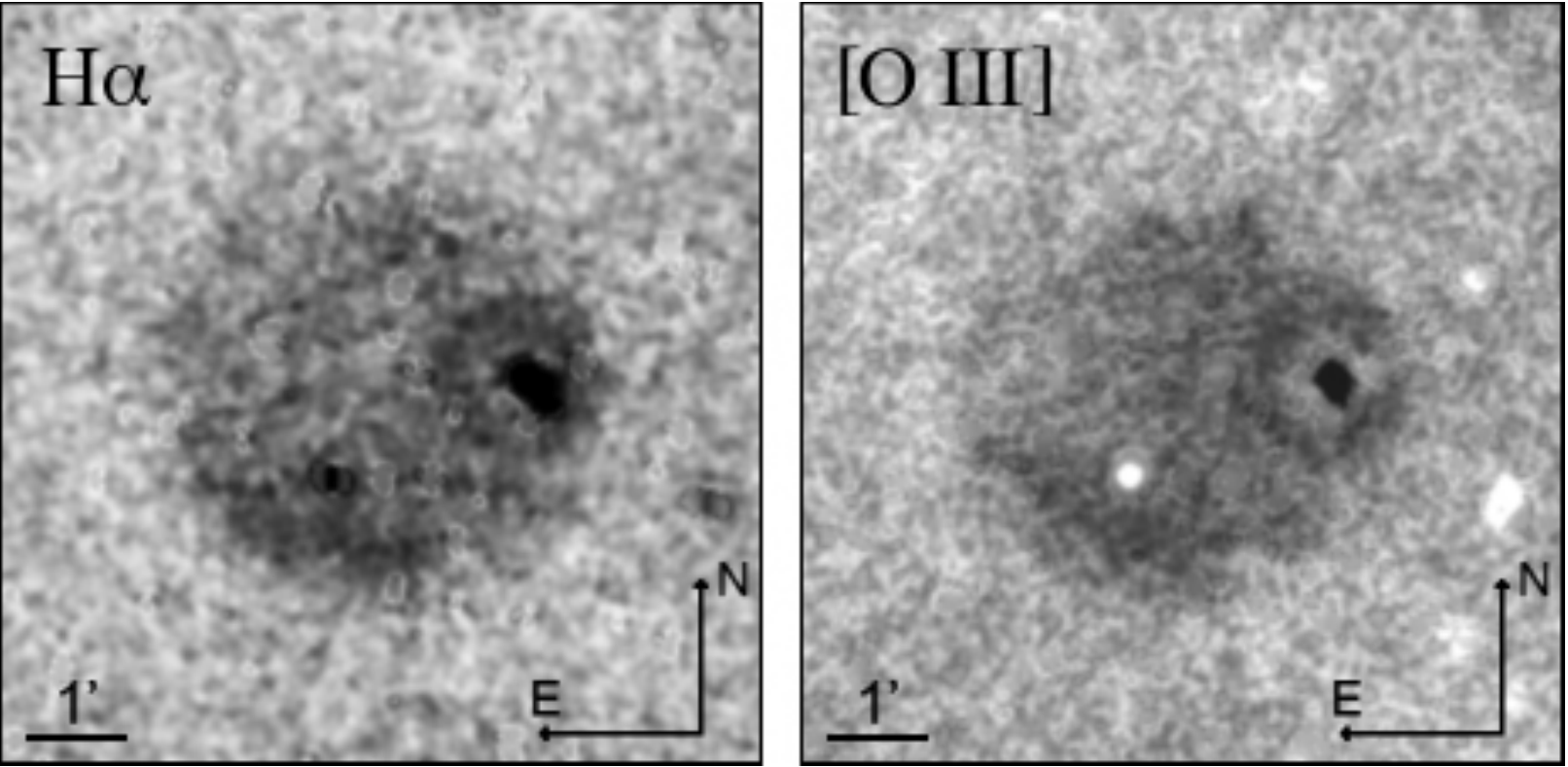}
     \includegraphics[scale=1.0,angle=0]{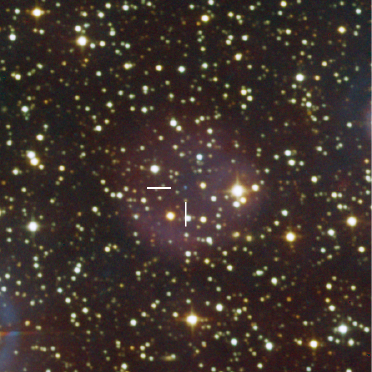}
  \caption{ Images  of Alv1 -- 
{\it Left:} H$\alpha$ and [O III] plates with continuum subtracted.
       {\it Right:} Location of the very blue candidate central star.}
  \label{fig3} \end{center}
\end{figure*}

Figure \ref{fig2} presents the reduced flux-calibrated spectra of Ou2 and Ou4. The spectrum of Ou4 was extracted by adding the brightest 25\arcsec\ of emission and subtracting an adjacent sky region of the same size. Due to the very low surface brightness only the strongest emission lines can be seen. 
Table~\ref{tab:spectro} gives the measured line intensities. In both cases the reddening is very low, but difficult to accurately measure.

These spectra were taken for classification purposes and therefore only the brightest nebular lines are measurable. The remaining apparent peaks are sky subtraction residuals: no emission is visible for the lines of [SII]-6717/32, and [NII]-6584 is just detectable for Ou2. This is a known situation : for most classical bright PN, the [SII]-doublet total intensity is very low (and can be about 150 times fainter than H$\alpha$ or not measurable), and a similar situation is observed for [NII]-6584 (see the line intensities of galactic PN in Acker et al. 1992). 
Note that in the case of shocked excited nebulae, the [SII]-lines could be as intense as H$\alpha$ and [OI]-6300 lines.

\section{Analysis of the 6 PN candidates}

For the four objects Alv1, lD\^u1, Ou2 and Ou3, their round morphology and blue CSPN are convincing evidence that they are fairly old PN. The object Ou4 presents an extended bipolar morphology and is of great angular extent. We discuss it extensively in Section 4.6. The Ou1 candidate is most likely an H~II region.

\begin{figure*}
  \begin{center}
    \includegraphics[scale=0.267]{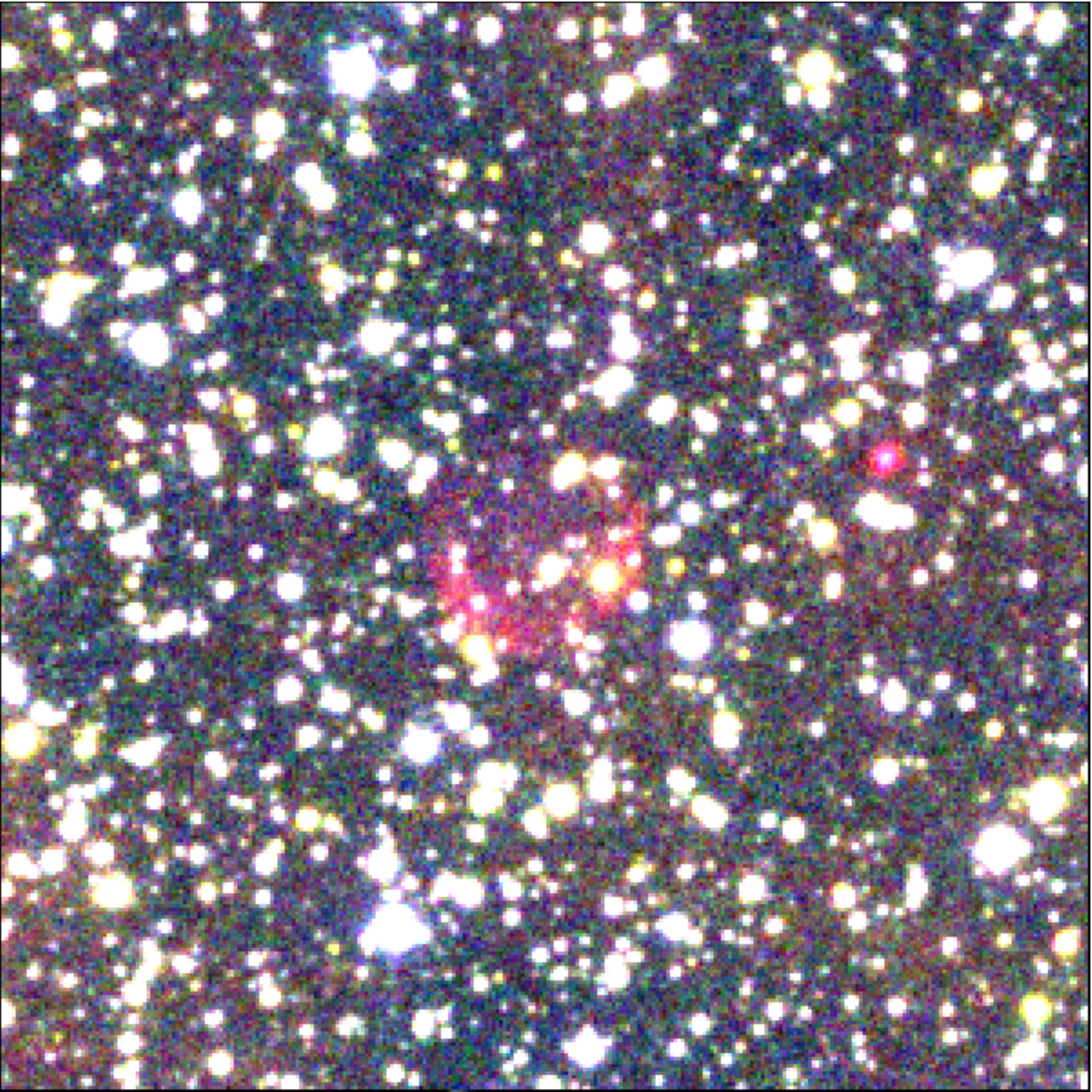}
     \includegraphics[scale=0.091,angle=0]{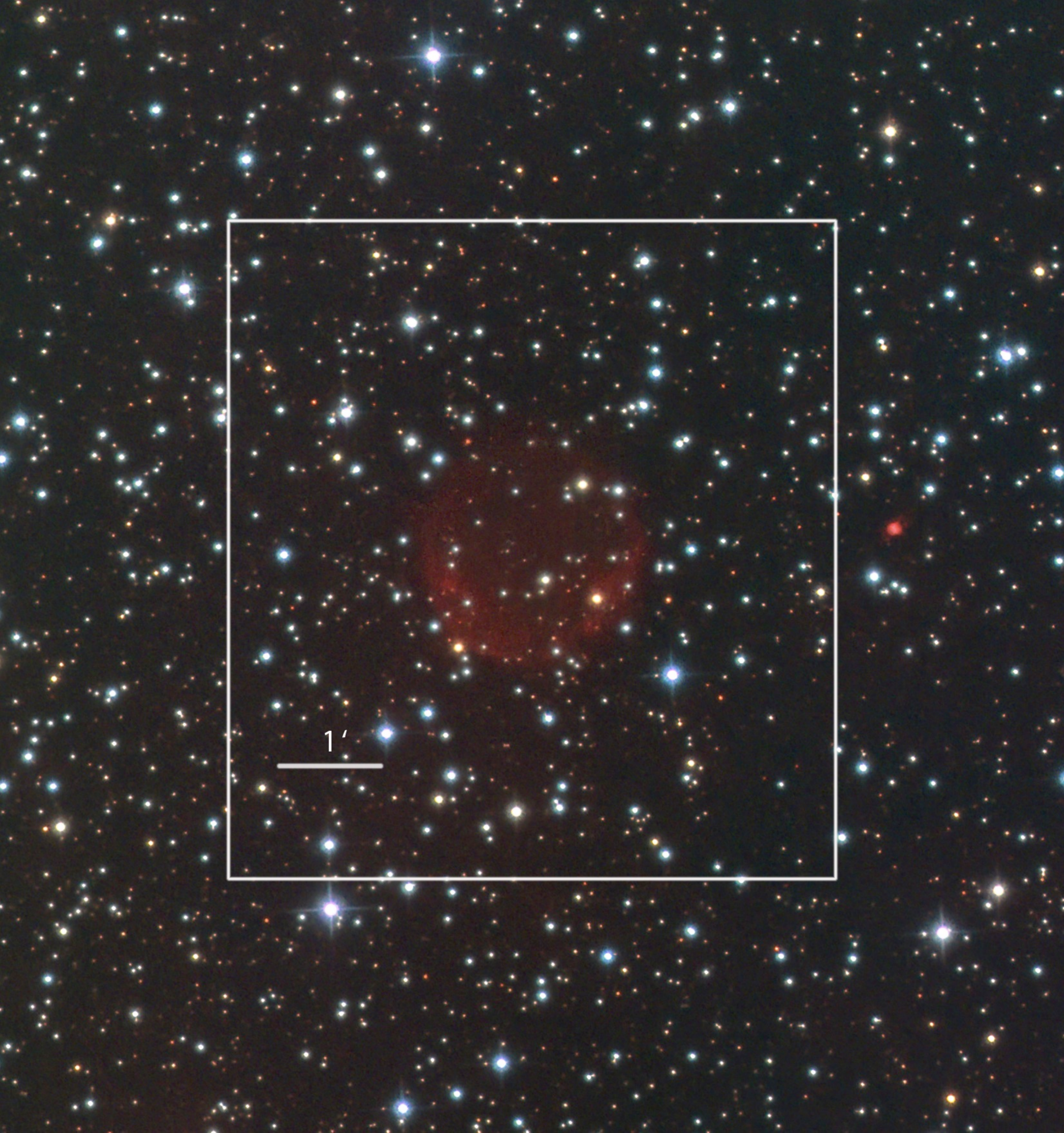}
     \includegraphics[scale=0.833,angle=0]{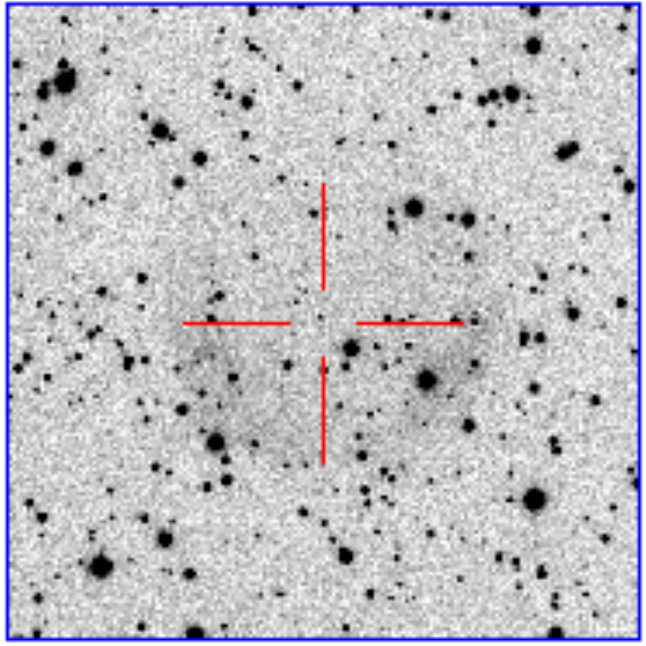}
  \caption{{\it Top left:} Zoom onto lD\^u~1, together with the nearby (208\arcsec) known compact PN K~3-83 (PN G094.5$-$00.8). The $10.25\arcmin\times10.25\arcmin$ image is made from H$\alpha$ (red), [S~II] (green) and [O~III] (blue). {\it  Top right:} zoom on lD\^u1, with the small PN K3-83 at right  (image H$\alpha$RGB composite, by Josef P\"opsel, Stefan Binnewies, Capella Observatory -- 9h exposure). 
{\it Bottom:} H$\alpha$+[N~II] image of the PN from the IPHAS survey  ($150\times150$ arcsec$^2$ and 1.33\arcsec/pixel). North is up, East to left in all images.}
  \label{fig4}  \end{center}
\end{figure*}

\subsection{Alv~1 -- PN G079.8-10.2}

A faint PN candidate was serendipitously discovered during an imaging session of the exotic planetary nebula MWP 1 (PN G080.3-10.4) with a wide field of 60\arcmin ~by 80\arcmin\ centered on MWP1.
Optical observations were performed remotely at the CROW observatory in Portalegre, Portugal, during several observing runs. Images were taken using a 60\AA ~wide H$\alpha$ filter and a 120\AA ~wide [O III] filter, as well as wide-band $BVR$ images. Exposure times ranged from 600s to 1800s using 1x1 or 2x2 binning depending on seeing and weather conditions. Exposures were later properly calibrated (i.e. debiased and flat-fielded), aligned and co-added.

Images taken during about 23 hours with the wide-band Astronomik $R$ and $V$ filters were used to subtract the continuum from the H$\alpha$ and [O~III] plates respectively; convolutions were applied in order to homogenize the PSFs between plates, and subtraction weights were found empirically to remove the stars. Fig. \ref{fig3} (left) shows the final images obtained for Alv1.

 By overlaying the broadband $BVR$ plates on our image, a very blue star (Fig.~\ref{fig3} right) of $B$$\sim$18.2 is found approximately 12\arcsec\ away from the geometrical centre of the nebula, and could possibly be the central star of this object, thereby increasing the possibility of it being a PN. 
We note that it is not unusual to find a slightly offset central star: this could be related to ISM-wind interaction, high proper motion of the central star, the presence of a binary system at the centre, or simply some asymmetry of the nebula itself. An extensive kinematical study would be necessary to assess the reason for this, but such a study is beyond the scope of the current paper.

\subsection{lD\^u 1 -- PN G094.5-00.8a}

 Three small round nebulae were found  in a 90\arcmin$\times$50\arcmin\ field around the H~II region Sh2--124 in July and August 2011. Two of them are known PNe, and the other is a much fainter new nebula (Fig.~\ref{fig4}). 43.3 hours of observations with H$\alpha$, [S~II], and [O~III] filters were accumulated to confirm that this is a new PN with a very faint central star, also seen on previous images from the IPHAS Survey (Fig.~\ref{fig4} bottom).

\begin{figure}
  \begin{center}
    \includegraphics[scale=0.45,angle=0]{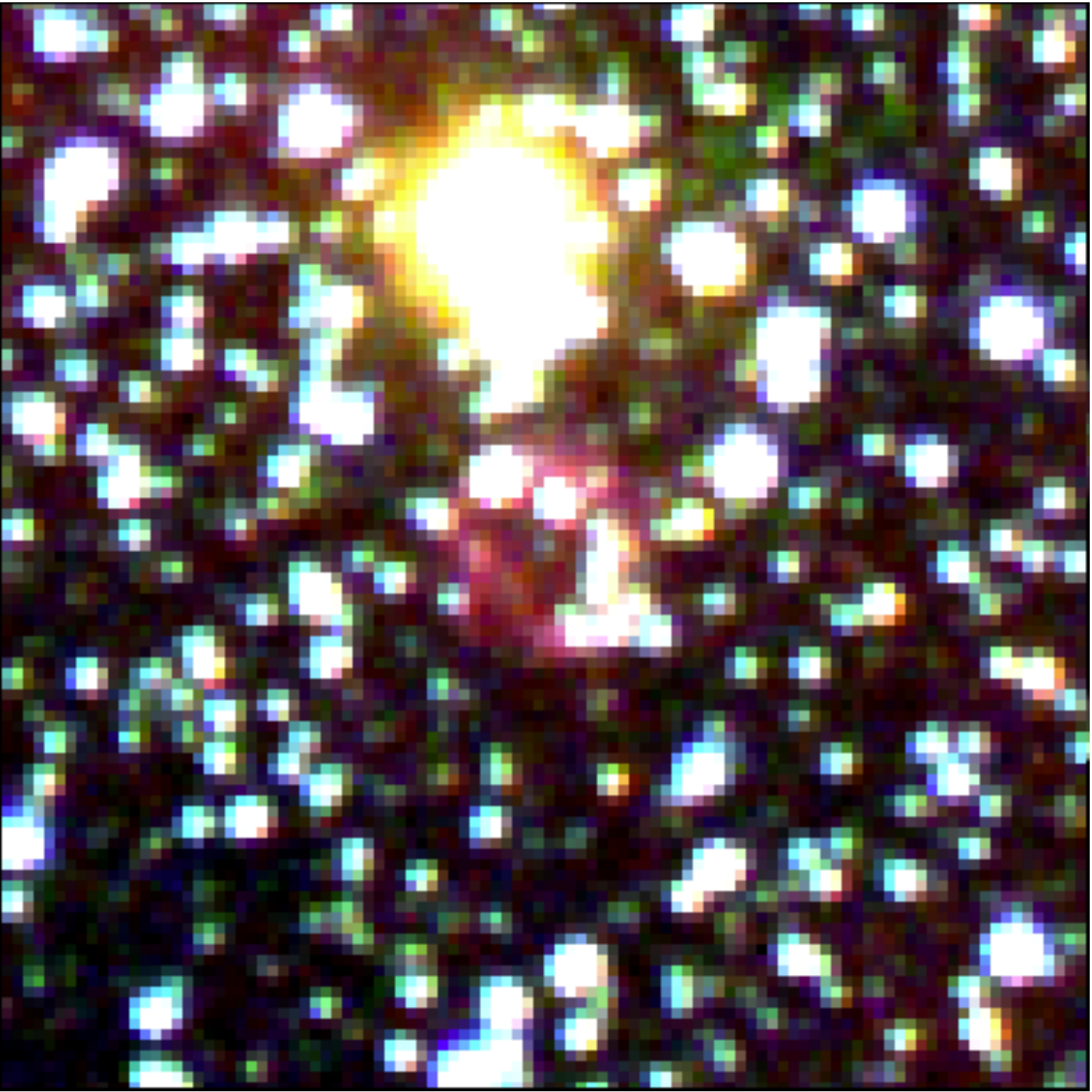}\\
     \includegraphics[scale=0.25,angle=0]{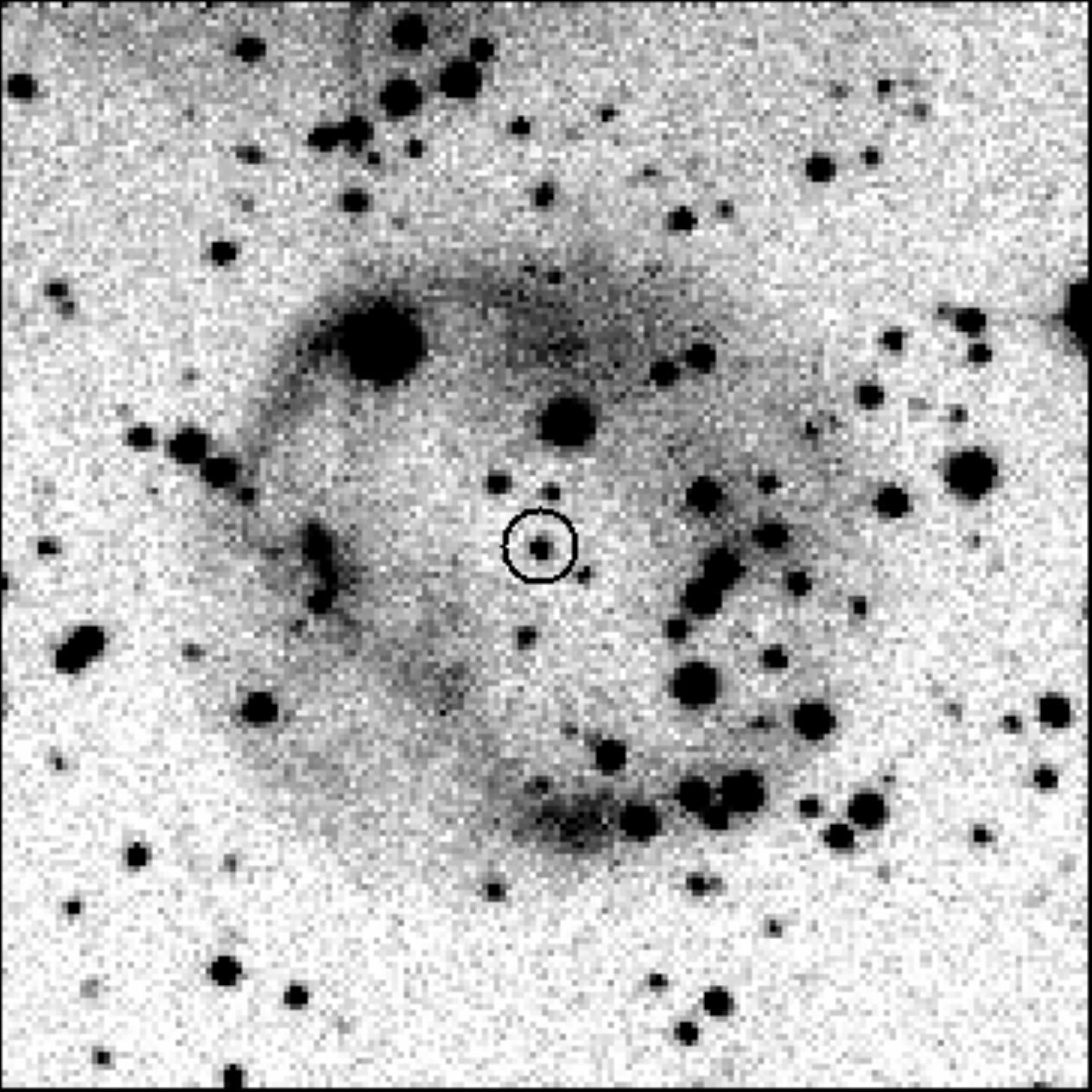}
     \includegraphics[scale=0.175,angle=0]{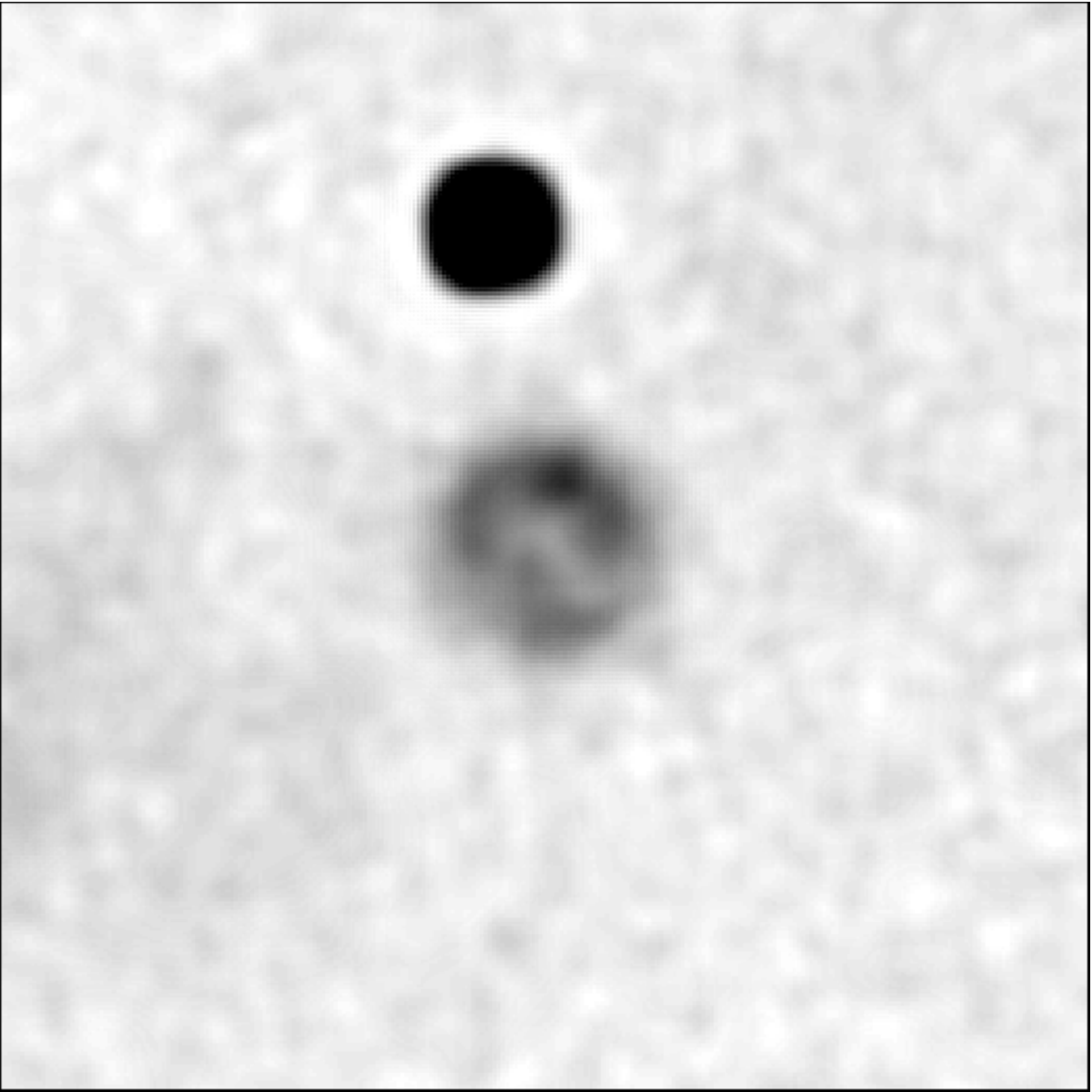}
  \end{center}
  \caption{ {\it Top :}  composite H$\alpha$RGB image showing the new PN Ou2 ($6\times6$ arcmin$^2$).
{\it Bottom left:} H$\alpha$+[N~II] image from the IPHAS Survey ($120\times120$ arcsec$^2$) with the central star candidate circled.  {\it Bottom right:}  \emph{WISE} $W4$ (22 $\mu$m) image ($6\times6$ arcmin$^2$) that has had a small amount of unsharp masking applied. North up, East left.}
  \label{fig6}
\end{figure}

\begin{figure*}[bt]
  \begin{center}
     \includegraphics[scale=0.19,angle=0]{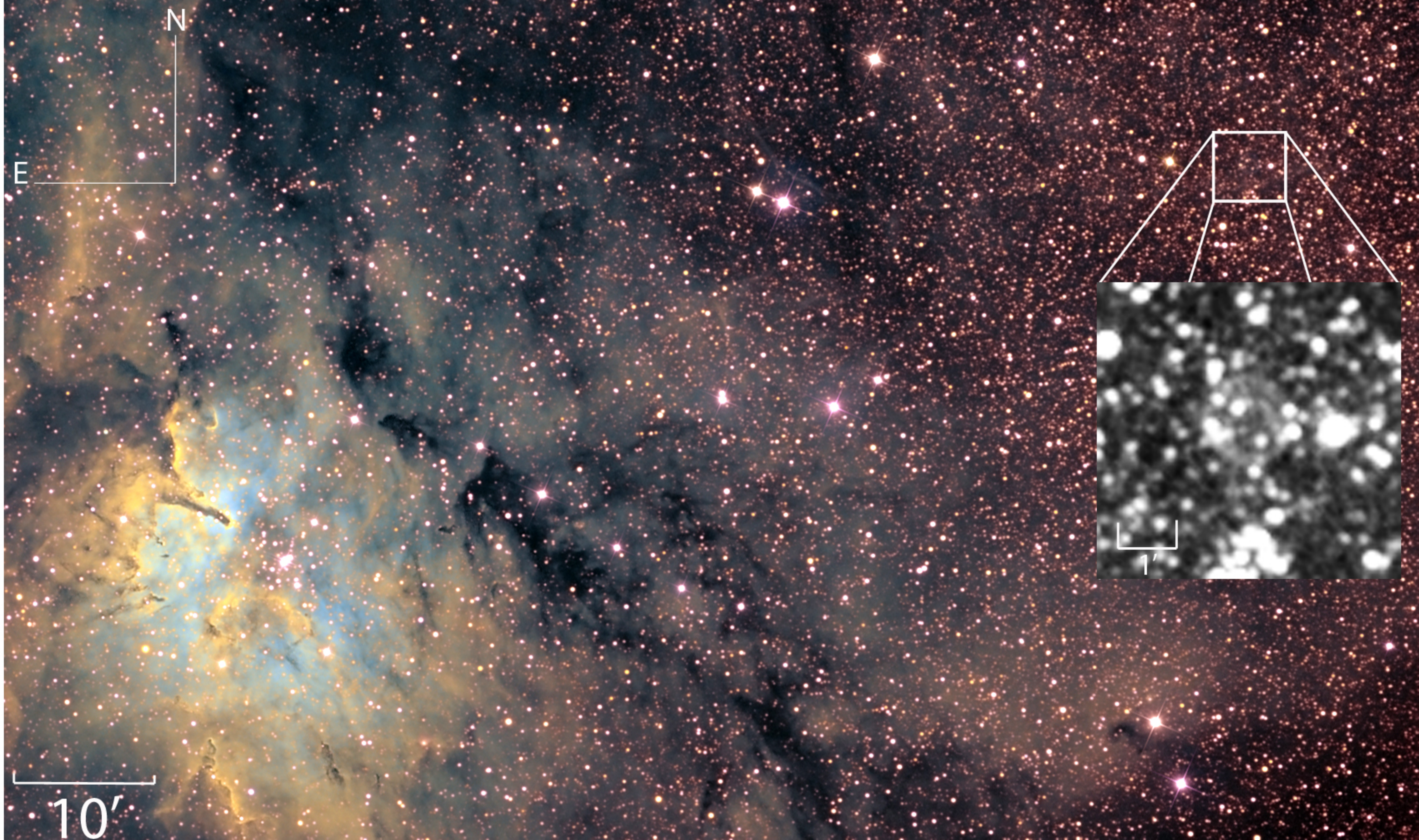}
  \end{center}
  \caption{Field around the nebula NGC 6820 (composite H$\alpha$RBV), with the object Ou3 shown in [O~III] in the 6\arcmin$\times$6\arcmin\ insert. 
}
  \label{fig7}
\end{figure*}

\begin{figure}[bt]
  \begin{center}
     \includegraphics[scale=0.52,angle=0]{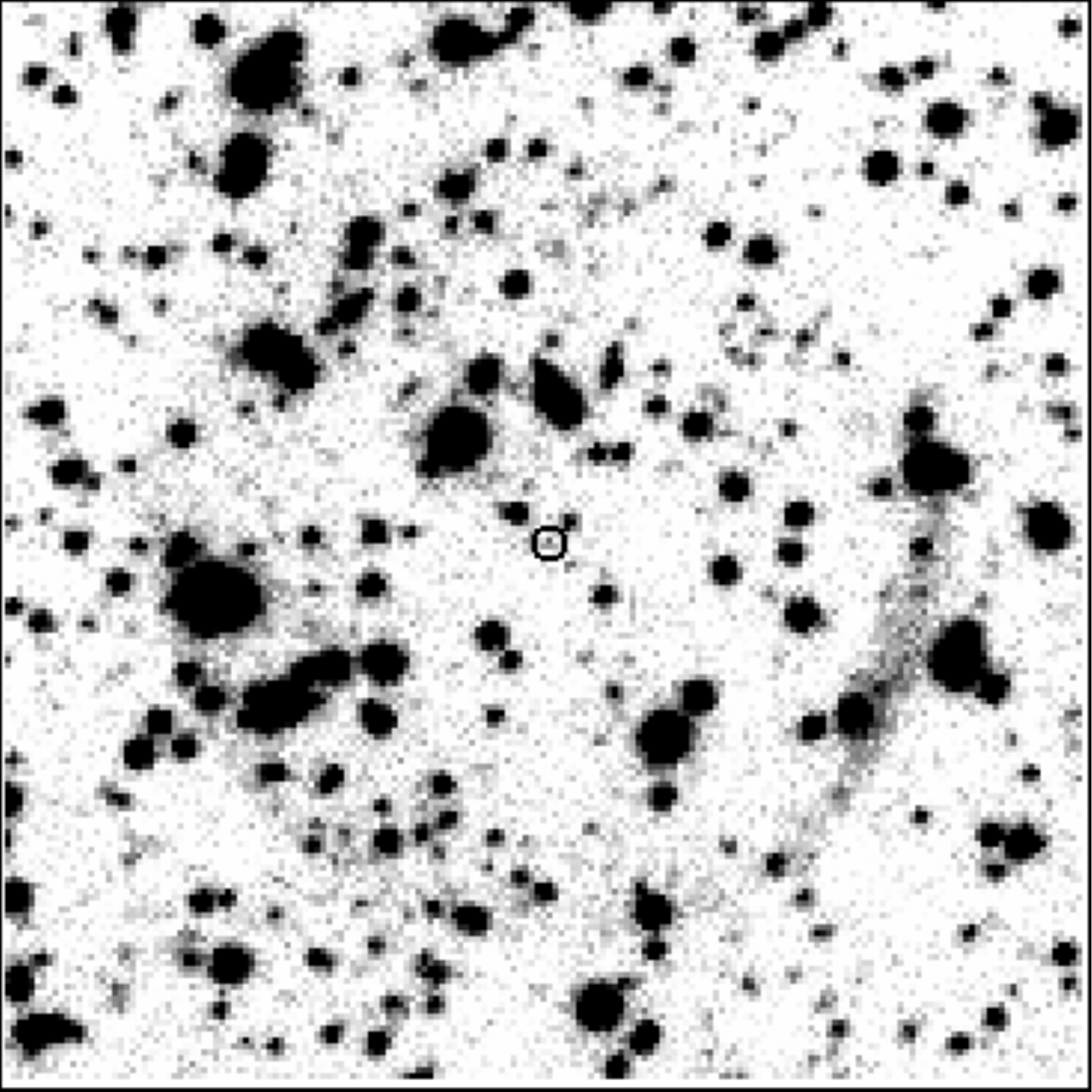}
  \end{center}
  \caption{H$\alpha$+[N~II] image of Ou3 from the IPHAS Survey (North up, East left).}
  \label{fig7b}
\end{figure}

\subsection{Ou2  -- PN G120.4-01.3 }

Accumulating 41 hours of observations in H$\alpha$ of the field around the H~II region Sh2--173,  a small round object (1.5\arcmin~in diameter) was discovered, with a central star appearing very blue on the $B$ image. The PN is visible on previous IPHAS images (Fig.~\ref{fig6}).

The spectrum (Fig.~\ref{fig2}) is dominated by the green [O~III] doublet and the high intensity of the He~II 4686 line (see Table~\ref{tab:spectro}). This implies a very high excitation, and a very hot CSPN.

\subsection{Ou3 -- PN G059.2+01.0}
This object was discovered on a narrow-band image of the field around the HII region NGC 6820, after exposing for 36 hours. As shown on Fig.~\ref{fig7}, a double ring is well detected on the [O~III] image,  with a diameter of about 1.8\arcmin.  A very faint nebula around a possible CSPN appears on the corresponding IPHAS image (Fig.~\ref{fig7b}).

\begin{figure}
  \begin{center}
    \includegraphics[scale=0.73,angle=0]{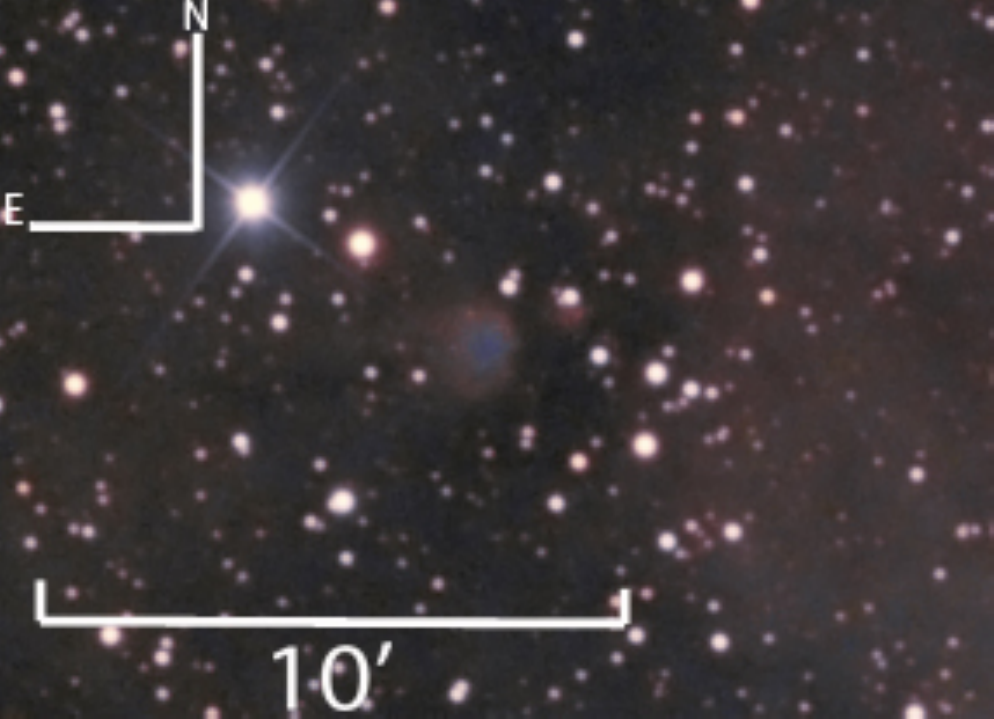}\\
     \includegraphics[scale=0.29,angle=0]{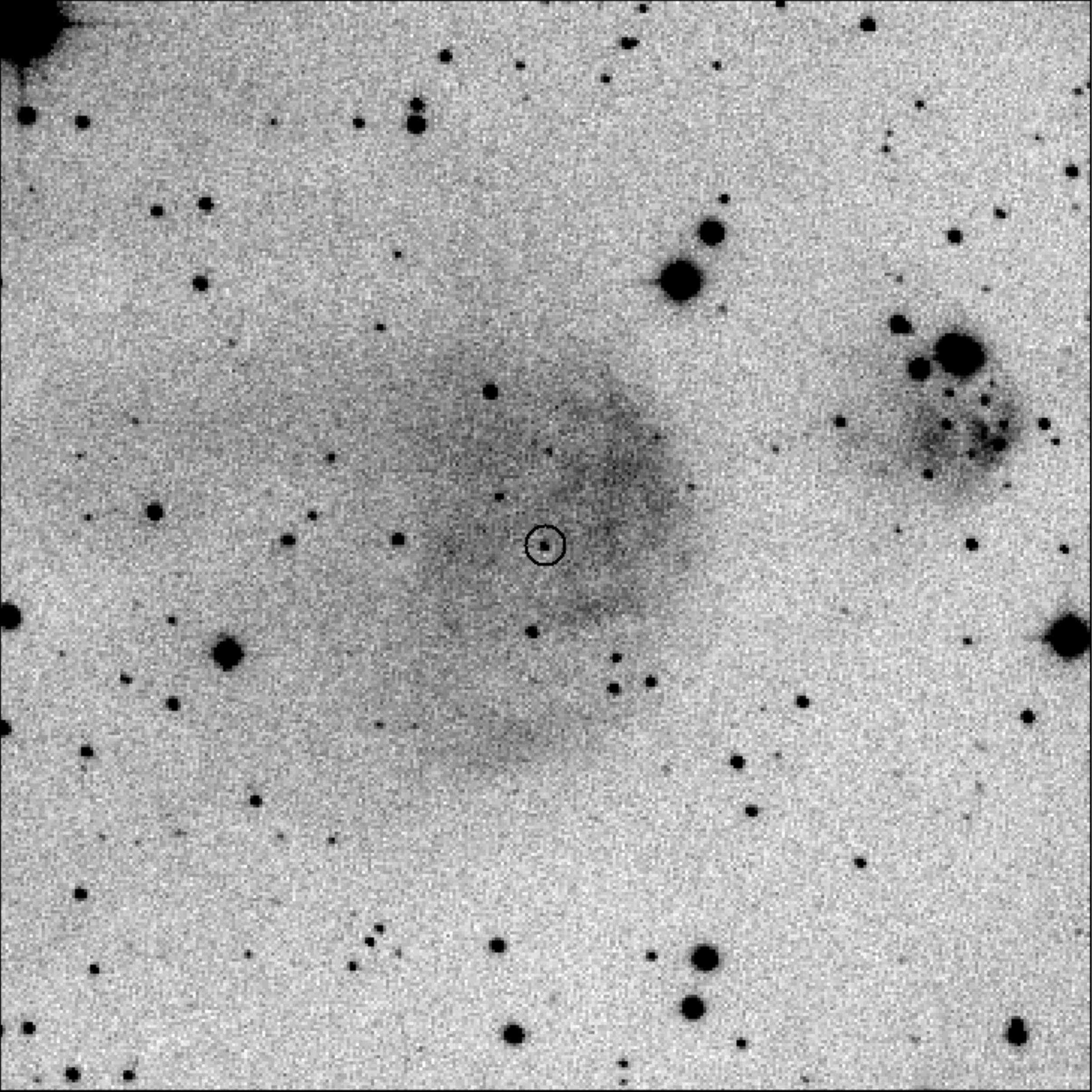}
  \end{center}
  \caption{
{\it Top:} Discovery colour-composite image of Ou1 in [S~II], H$\alpha$, and [O~III]. 
{\it Bottom:} H$\alpha$+[N~II] image from the IPHAS survey.The circle indicates the position of the likely central star.}
  \label{fig5}
\end{figure}

\begin{figure*}[bt]
  \begin{center}
     \includegraphics[scale=0.25,angle=0]{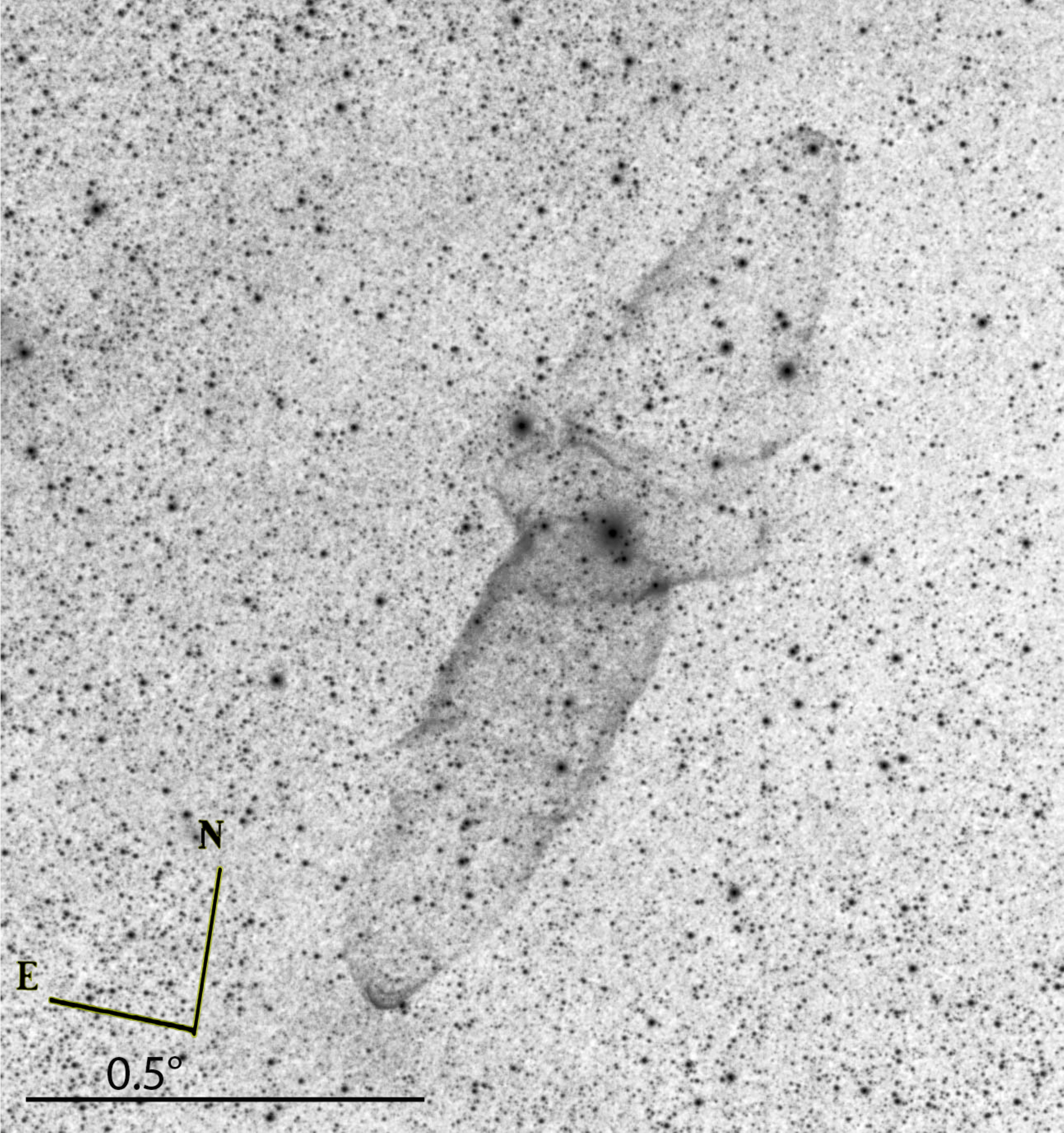}
  \end{center}
  \caption{
Ou4 image obtained by 25 exposures of 30 min [O~III]-3 nm filter.
}
  \label{fig8}
\end{figure*}

\subsection{Ou1 -- A non-PN but H~II region?}
Studying in detail the region surrounding the H~II region NGC~1491,  a faint nebula of about $1.2\arcmin\times1\arcmin$ extension was discovered in the field. Exposures of 10-min each were accumulated over 4 consecutive nights for a total of 24 hours of observations with Astrodon 6~$\mu$m-wide H$\alpha$, [S~II] and [O~III] filters. The final image was composed using the {\tt Ccdautopilot3} software (Fig.~\ref{fig5}). A very blue star lies roughly at the centre of the nebula and is certainly responsible for the observed nebula seen in the [O~III] image. 

The asymmetric morphology of Ou1 bears strong resemblance to that of A35, a PN distorted by interaction with the interstellar medium due to its high proper motion \citep{AFPJ98}.
But in Ou1, the star is very close to the centre and not located near the compressed zone. Given its very close proximity to another H II region and its patchy
H$\alpha$ morphology unlike other PNe, we classify Ou1 as a likely H~II region rather than a PN. Ultimately a spectrum of the $B_J=19.38$ mag ionising star will decide between the two possibilities.

\subsection{Ou4  -- PN G 098.6+07.9 (TBC)}
While observing Sh2--129, a strange object was discovered inside this large nebula: with the narrow [O~III] filter, a bipolar nebula appears (Fig.~\ref{fig8}). Its shape is similar to well-known bipolar PNe, such as Mz3 \citep{SG2004}, M2--9 \citep{Cor2011}, and KjPN8 \citep{Lo2000}. 

 The morphology appears as a complex addition of two components: a small central disc-like structure and two elongated opposite lobes -- comparable to the large faint ``chakram'' and the relatively short lobes of Mz3, both being seen in H$\alpha$+[N~II] light \citep[Fig.~1]{SG2004}. 


The spectrum of Ou4 (Fig. \ref{fig2} and Table~\ref{tab:spectro}) is dominated by the green [O~III] doublet, whose total intensity is about 20 times larger than that of H$\beta$. Such a high excitation class is coherent with a very hot CSPN.


The very bright star lying in the central part of the nebula is HR 8119  ($\alpha=$21 11 48.18, $\delta=$+59 59 09), of spectral type B0~II or B0~V, depending on the authors (see below). 
The star is thought to belong to the Cepheus OB2 massive stars association, the Sh2--129 H~II region, or the Trumpler 37 cluster, and is variable  (V=5.51 to 5.57). The star belongs to a triple system 
%
%
 \citep{EP08}, 
with the following configuration in magnitude-spectrum and separation for the brightest components: Aa: 6.17-B0~II + Ab:~? ; separation of 0.06\arcsec, and B: 6.78; separation AB of 1.035\arcsec. 
Using speckle interferometry, \citet{M2009} derived the following parameters for the triple system: 
\begin{itemize}
\item stars AaAb: position angle = 124.5$^{\circ}$, separation = 0.045\arcsec; $\Delta V = 0.6$
\item stars AB: position angle = 212.5$^{\circ}$, separation = 1.013\arcsec, $\Delta V = 0.3$.
\end{itemize}

We may thus attribute the following magnitude values for the components: star Aa:~6.17,  star Ab:~6.17+0.6~$\sim$6.8;
star B:~6.17+0.3~$\sim$6.5. Could the unknown faint component Ab of the B0~II star be a CSPN? We consider this and the opposite in the following.

We have also to take into account another characteristics of the system of HR 8119: the ROSAT all-sky survey includes the source 1RX J211148.9+595920, at about the same position than HR 8119 with a 14\arcsec ~uncertainty. The source is found to have a flux of about $10^{-13}$ erg/s/cm$^2$ and a count rate of 0.033$\pm$0.008 counts/s.

\vskip 6pt
\paragraph {\it a) Is Ou4 connected with HR 8119 ?}
Using the relation between the strengths of the interstellar Ca lines in the, assumed, B0~II star spectrum, and the distances to early-type stars, \citet{Me2009} estimated a distance of 1032$\pm$144 pc, in agreement with the (poorly determined) parallax measured by Hipparcos (1315$\pm$726 pc). \citet{TNH2011} consider on the other hand that HR 8119 is a 37.2 M$_\odot$ young B0~V star, located in the Trumpler 37 cluster, at a distance of 835 pc. 
If we adopt a realistic absolute magnitude for such a B0~V star, i.e. M$_{\rm V}=-3.6$  \citep{Wal1972}, with an apparent magnitude of 6.1 (component Aa), we would find a distance of $\sim$870 pc, compatible with the result of  \citet{TNH2011}. Note that the H~II region Sh2--129 is closer than HR~8119: its kinematic distance is estimated as 400$\pm$130 pc \citep{BB1993}.

We examine different possibilities based on the association of Ou4 and HR 8119.

If we assume that a PN emerges from the physical companion to HR~8119, we would of course be faced with 2 major problems. First, the young age of the system would preclude the evolution of a star up to the PN stage. And obviously a massive star cannot be a CSPN. Second, 
the large angular dimension (1$^{\circ}$10\arcmin33\arcsec $\times$ 19\arcmin54\arcsec) would lead to an extension of the nebula of about 20 pc  at a distance of about 1 kpc -- much too large for a PN! As a comparison, the bipolar PN Mz3 (covering 1.5\arcmin on the sky) extends only 0.5--1.2 pc, depending on the estimated distance (1 to 2.5 kpc; \citet{SG2004}).


Could Ou4 be a shell recently ejected by a nova associated with the X-ray source, as was for example seen for V445 Puppis \citep{Woudt}?
A priori, such a X-ray source could be associated with a close binary with an accreting white dwarf in the centre of Ou4, in agreement with the high excitation class of the nebular spectrum. 
However, if the object was located at the distance of 800-1000 pc, its intrinsic luminosity would be too high for a cataclysmic variable. We would thus have to assume it is even further away, which makes it then difficult to reconcile with the extent of the nebula on the sky. 

Another possibility is that Ou4 is some other kind of nebulosity related to HR~8119. Recently, \citet{Ange2011}
discovered a giant, highly collimated jet from Sanduleak's star in the Large Magellanic Cloud, with a physical extent of 14 pc. Sanduleak's star shows an  optical emission-line spectrum reminiscent of a dusty-type symbiotic star, including the presence of the Raman bands at $\lambda\lambda$6825, 7082.  Clearly this is not the case for HR~8119. 
However, its mere existence proves that large outflows of such nature are possible.

It seems that Ou4 is probably not associated with the distant B0 star HR 8119.

\vskip 6pt
\paragraph {\it b) Is Ou4 a near PN?}
 One intriguing possibility would be that it is a highly excited PN, with a very hot CSPN, seen in the same direction as HR 8119, and, hence, seen as an optical companion (wrongly identified as Ab thus -- see above), but without any physical link with the B0 star. 
A distance could be estimated by comparison with Mz 3: a typical extent for such PN would be 0.5--1 pc, which given the angular size of Ou4 would translate into a distance of $\sim$25--50 pc.  We note that the distance seems to be relatively small because the blue part of the spectrum (Fig.~\ref{fig2} and Table~\ref{tab:spectro}) shows negligible extinction. 
If this were confirmed, Ou4 would be the nearest PN discovered to date!
At such a distance, the corresponding absolute magnitude of the optical companion Ab ($V$$\sim$6.8) to HR~8119 would be 3.5--4.9, which is compatible with typical values for CSPNe.

Note that detections of diffuse X-ray emission were reported for more than 20 PNe, in particular BD+30$^\circ$3639 (count rate 0.242$\pm$0.004 cnts s$^{-1}$, Flux 6.8$\times$10$^{-13}$ erg cm$^{-2}$ s$^{-1}$, Kastner et al. 2000), and Mz 3 (count rate 0.0022$\pm$0.0003 cnts s$^{-1}$, Flux 7.0$\times$10$^{-15}$ erg cm$^{-2}$ s$^{-1}$, Kastner et al. 2003), an emission associated here with the bow-shocks of the fast collimated outflows. As claimed by Guerrero et al. (2000), PNe are key to assessing the action of fast stellar winds and collimated outflows in the formation and evolution of PNe near the end of the AGB phase. 
However the X-ray source observed here is based on extremely shallow all-sky measurements and consistent clearly with the B0 type star HR 8119.

At this stage, we are thus led to conclude that the nature of Ou4 is controversial: it could be a massive collimated outflow but without relation with the B0 star as it would produce an extremely large nebula considering the distance to the star. Or it is a PN which is $\sim$0.5--1 pc across and $\sim$25--50 pc away, in agreement with the absence of extinction. Ou4 should certainly become the target of many studies to come.

\section{Discussion and Conclusions}

In this paper, we have presented 6 newly discovered nebulae, obtained through very deep narrow-band imaging by amateur astronomers. Four of them (Ou2, Ou3, lD\^u1 and Alv1) are round, fairly evolved PNe, and correspond to the missing population of such PNe that is thought to exist. Such PNe are most easily discovered by the patient and dedicated work of amateur astronomers than automatic surveys, most efficient at discovering small, compact objects. The example of lD\^u1, located just next to the small PN K3-83 discovered in 1972 by Kohoutek is very revealing in this respect. The detection of new PNe by experienced amateur astronomers will thus be an excellent support to improve the completeness of the Galactic planetary nebula census.

The fifth object is quite remarkable. Ou4 is a very large bipolar nebula, with a central ``chakram'', opposite tubular lobes emerging from an invisible, hot central star. 
 If confirmed as a PN, Ou4 may be the nearest PN known to date, with a possibly identified central star. If on the other hand, it is located at a distance of 800 pc to 1 kpc, it would be one of the largest bipolar nebulae discovered. Whatever its final nature, the complex shape and amazing properties make it an object of choice for further studies.

Most sources are located near the Galactic Plane where many star forming complexes, H~II regions, and molecular clouds are present, and there is a high probability of a candidate being nearby in projection, or physically associated, with one of those regions: it is the case of Ou4 (related to Sh2--129), and of Ou1 associated with a nearby H~II region.
The new PNe near the galactic plane are smaller (mean size of 110\arcsec\ for 3 PNe) than those at higher latitude (70\arcmin\ and 270\arcsec). This could be related to the decreasing ISM density, although any shock caused by the ISM would make the PN much brighter than they are observed.

N. Outters observed 49 fields, whose sizes vary between 24\arcmin\ $ \times $ 16\arcmin\ and 180\arcmin $\times $ 119\arcmin, and found three candidate PNe. P. LeD\^u found four PNe -- among which one new candidate -- after observing 11 fields, of sizes between 80\arcmin$\times$ 60\arcmin\ and 116\arcmin$\ \times$ 88\arcmin. Compared to the number of PNe already known in these regions of the sky, this amount to an increase of about 10\% in the number of PNe. This is still far from the factor 10 we are looking for, as predicted. 

We are, however, convinced that the contribution of amateur
astronomers to the discovery of relatively nearby PNe inside the Milky Way will  increase in the coming years. This could provide essential clues regarding the final number of PNe in the Galaxy and their origin. In particular, we stress the important role amateur astronomers could play in the less explored high galactic latitude regions, outside current professional H$\alpha$ surveys.

\acknowledgements
AA is grateful to Ch. Motch and M. Pakull from the Strasbourg Observatory for their help in the identification of the X-ray counterpart of Ou4. Many thanks to Josef P\"opsel and S. Binnewies ({\tt www.capella-observatory.com}) for providing the superb image of lD\^u1. 
FA would like to thank the CROW team (Jos\'e Canela, Jorge Canela and Paulo Barros), the Atalaia team (Alberto Fernando, Jos\'e Ribeiro and Joa\~o Greg\'orio and Lic\'\i nio Almeida) for their valuable support, and Phillippe Stee, George Jacoby, Lubos Kohoutek for their useful advice. LS is supported by PAPIIT-UNAM grant IN109509 (Mexico). BM gratefully acknowledges the ESO Chile Visitor Programme.


\end{document}